\documentclass[aps]{revtex4-1}
\usepackage{graphicx}
\usepackage{natbib}
\usepackage{bm}
\usepackage{amssymb}
\usepackage{amsmath}
\usepackage{euscript}
\usepackage{esint}
\usepackage{slashed}

\begin{document}

\title{Analytical study in the mechanism of flame movement in horizontal tubes. \\ II. Flame acceleration in smooth open tubes}

\author{Kirill A. Kazakov}

\affiliation{Department of Theoretical Physics, Physics Faculty, Moscow State
University, 119991, Moscow, Russian Federation}

\begin{abstract}
The problem of spontaneous acceleration of premixed flames propagating in open horizontal tubes with smooth walls is revisited. It is proved that in long tubes, this process can be considered quasi-steady, and an equation for the flame front position is derived using the on-shell description. Numerical solutions of this equation are found which show that as in the case of uniform flame movement, there are two essentially different regimes of flame propagation. In the type~I regime, the flame speed and its acceleration are comparatively low, whereas the type~II regime is characterized by significant flame acceleration that rapidly increases as the flame travels along the tube. A detailed comparison of the obtained results with the experimental data on flame acceleration in methane-air mixtures is given. In particular, it is confirmed that flames propagating in near-stoichiometric mixtures and mixtures near the limits of inflammability belong to the types II and I, respectively, whereas flames in transient mixtures undergo transitions between the two regimes during their travel.
\end{abstract}
\maketitle

\section{Introduction}

It is well-known that burning of gaseous mixtures in tubes often proceeds in accelerated way -- the apparent propagation speed of the flame front increases with time. Observations show that the occurrence of this phenomenon and its main characteristics essentially depend on the mixture composition and the physical conditions of combustion -- boundary conditions at the tube ends (it can be open, semi-open or closed), location of the ignition point, tube wall roughness, presence of acoustic fields, and so on. Although precise conditions leading to the flame acceleration have not been identified yet, it is generally agreed that importance of each of the above factors is determined by the extent to which it perturbs the flame. This view is naturally based on the fact that flame perturbations increase the flame front area, and hence the overall gas consumption rate. In the studies of deflagration-to-detonation transitions, for instance, mixture is usually ignited near the closed end of a tube with roughed walls and/or equipped with a series of restricting rings.\cite{chapman1926,shchelkin} Because of the high irregularity of flame movement under such conditions, no quantitative description of this {\it stimulated} flame acceleration exists, though its mechanism is qualitatively understood fairly well.

\begin{figure}
\centering
\includegraphics[width=0.75\textwidth]{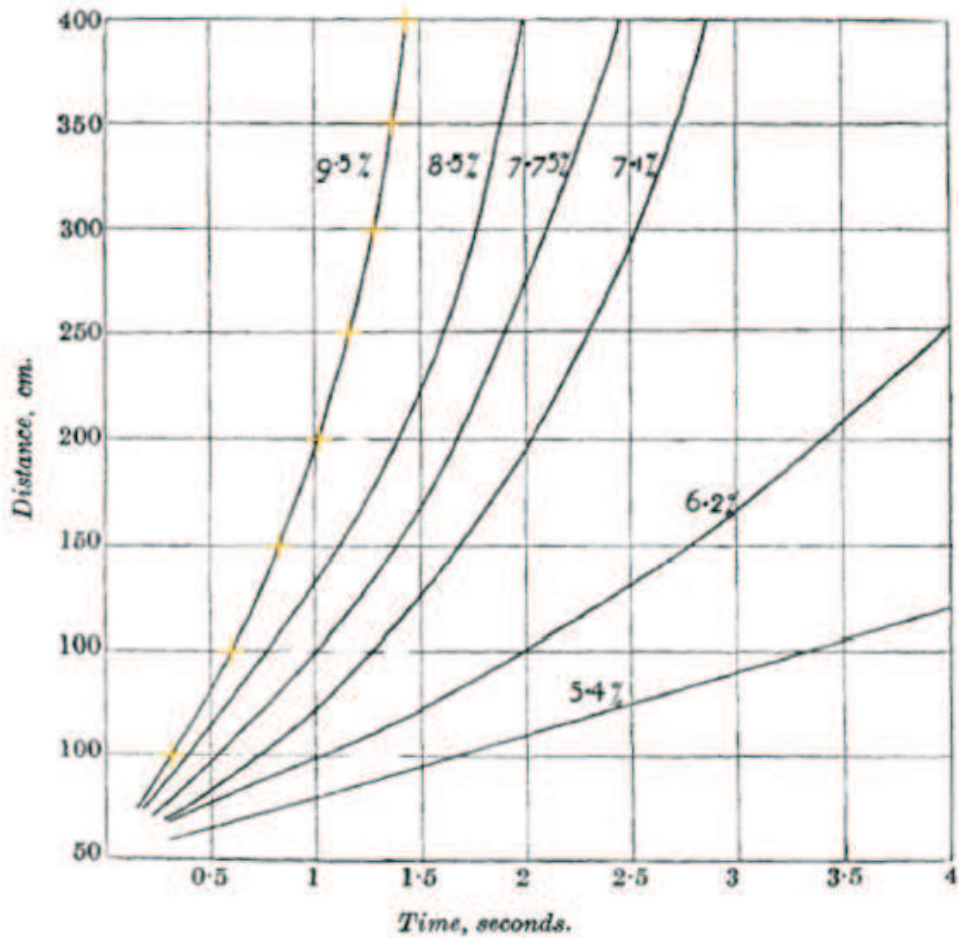}
\caption{Trajectories of flames propagating in various methane-air mixtures. The marks on the $9.5\%$ curve are the basis points used to construct an approximating polynomial (Sec.~\ref{processing}). {\it Source}: Fig.~1 of Ref.~\cite{mason1920a} -- Reproduced by permission of The Royal Society of Chemistry.}\label{fig1}
\end{figure}

On the contrary, flame acceleration in tubes with smooth walls, or {\it spontaneous} acceleration, though also known well, is understood much less.\cite{footnote1} Historically, it was discovered and studied\cite{mason1920a} earlier than the stimulated flame acceleration, as it exists in most mixtures burned in tubes open at both ends. Unlike its counterpart, spontaneous acceleration is a quite regular phenomenon, in that it takes place in laminar regimes of flame propagation. Moreover, experiments demonstrate reproducible flame behavior during acceleration which is determined by the tube size and mixture composition. Figure~\ref{fig1} shows experimental plots of the flame position as a function of time for flames propagating in an open tube $5\,$cm in diameter and $5\,$m long, for various methane-air mixtures.\cite{mason1920a} A striking evidence contained in the figure is the difference in the values of flame acceleration in different mixtures. Even without detailed analysis, it is clear that the rate of change of the tangent to the flame trajectory in the stoichiometric mixture ($9.5\%$ methane) is much higher than that in the lean mixtures. Moreover, a certain grouping of curves is evident: trajectories in mixtures $7.1\%$ to $9.5\%$ are quite similar, and are clearly separated from the $6.2\%$ and $5.4\%$ curves. The following reasoning makes the apparent peculiarity of these observations more concrete. Dynamics of flames under consideration is controlled by gravity: the dimensionless gravitational acceleration,
$$g = (acceleration~of~gravity)\times  (tube~diameter)/(normal~flame~speed)^2$$ exceeds unity for any methane concentration. The conclusion commonly made in this case is that the flame propagation speed is independent of the normal flame speed (for instance, this premise is the basis of the ``bubble'' models\cite{shtemler1996,bychkov1997,bychkov2000}). Therefore, it would be natural to expect universal behavior of all flames, in particular, similarity of flame acceleration, whatever specific mechanism of flame movement. We have just seen, however, that this contradicts observations. One might think that the observed differences could be ascribed to the differences in thermal expansion of gases, but invoking numerics readily rules out this possibility. To be specific, the gas expansion coefficients of the $6.2\%$ and $7.1\%$ mixtures differ by about $10$ percent, and this is also true of the $7.1\%$ and $7.75\%$ mixtures, but is clearly incompatible with Fig.~\ref{fig1}.

Even in tubes of such a moderate diameter as $5\,$cm, the speed of flame propagation largely exceeds the normal speed, and significantly increases as the result of acceleration. For instance, the normal flame speed in a $7.1\%$ mixture is $22\,$cm/s, whereas its propagation speed in the $5\,$m long tube is initially about $50\,$cm/s, increasing eventually to $350\,$cm/s. This means that spontaneous flame acceleration is a highly nonlinear process: the nonlinearity of gasdynamic equations describing the gas flows induced by the propagating flame in no approximation can be considered weak. This is why until quite recently this process had been beyond the reach of theoretical analysis (evidently, the conventional methods based on explicit solving of the flow equations are useless in problems of this sort). Things changed with the invention of the on-shell description of flames.\cite{kazakov1,kazakov2} This description will be used below to study the phenomenon of spontaneous flame acceleration in open horizontal tubes. It will be shown that the resolution of the above-mentioned peculiarity is in the existence of two distinct regimes of flame propagation, characterized by significantly different acceleration. The existence of two different regimes was first established in the study\cite{kazakov3} of the uniform flame movement in horizontal tubes (observed experimentally in semi-open tubes when the mixture is ignited near the open end), where it was noticed that they also differ regarding flame response to the heat losses: flames propagating in the slower type~I regime decelerate when the losses increase, while the faster type~II flames accelerate. It turns out that things are similar in tubes open at both ends: the two types of flames, identified in Ref.~\cite{kazakov3}, respond differently to the energy losses associated with the cold gas movement allowed by unflanging the tube, though now acceleration is positive in both regimes.

In view of the equivalence of gravity and translatory acceleration, reduction of the master equation to the ordinary differential equations for the flame front position and the on-shell gas velocity almost literally repeats that given in Ref.~\cite{kazakov3}, to be referred to below as Part~I. Therefore, it will not be reproduced here, and only the final result will be written down in Sec.~\ref{onshellequations}. Neither will be reproduced other important considerations such as derivation of the formula for the heat losses and the accuracy estimate, which remain the same, except the issue of identification of physical solutions. The corresponding criterion is now different because of the different conditions at the tube ends; its derivation will be given at full length in Sec.~\ref{ident}. Detailed comparison with the experiment is carried out in Sec.~\ref{comparison}. The concluding Sec.~\ref{conclusions} discusses the results obtained.

\section{On-shell description of accelerating flames}

\subsection{Physical conditions}\label{formulation}

\begin{figure}
\centering
\includegraphics[width=1\textwidth]{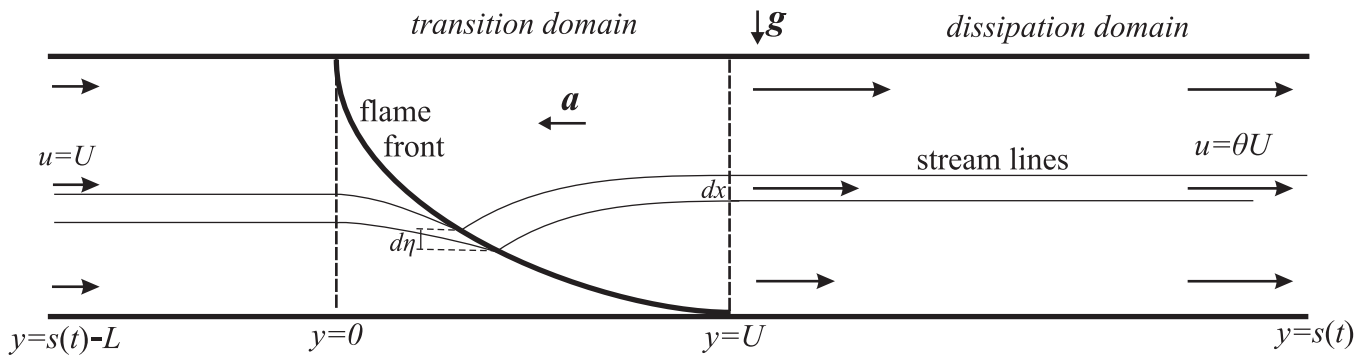}
\caption{Schematics of a flame accelerating in horizontal channel. The $y$-coordinate is measured in the natural units (Sec.~\ref{onshellequations}). The two vertical broken lines bound the transition domain $y\in(0,U)$ (Sec.~\ref{ident}). The pair of neighboring streamlines and the transversal line elements are used in writing Eqs.~(\ref{momentumup})--(\ref{bernoullidown}).}\label{fig2}
\end{figure}

Consider flame propagation in an initially quiescent gaseous mixture filling horizontal tube of diameter $d,$ open at both ends. Experiments show that soon after ignition at one end, the flame front assumes a characteristic shape depicted schematically in Fig.~\ref{fig2}. The less the normal flame speed, the stronger the front stretches along the tube. The subsequent flame evolution depends on the mixture composition and the aspect ratio of the tube. As evidenced by Fig.~\ref{fig1}, the flame may or may not accelerate, but an important observational fact is that the flame propagates quasi-steadily in any case, provided that the tube length $L$ is large enough, $L/d\gtrsim 50.$ More precisely, observations indicate that under this condition, the flame shape changes fairly slowly during most of its travel, except for relatively short periods of irregular changes during which the flame appears disturbed. In view of this, the phenomenon of spontaneous flame acceleration will be considered below in the quasi-steady approximation, defined formally as follows. At each time instant, we introduce a noninertial frame of reference moving with respect to the laboratory frame with the speed and acceleration of the flame at that instant. Then it is required that the flame be steady in the noninertial frame. Equivalently, in the laboratory frame, the quasi-steady condition requires the time derivatives of the flame speed of orders higher than the first be negligible. In other words, it is assumed that on every small interval the flame moves as if its acceleration were constant, equal to the true instantaneous acceleration on that interval. An analytic expression of this condition will be obtained in Sec.~\ref{quasisteady}. As a result of the transition to the noninertial frame, the acceleration is replaced by a uniform gravity field equal in value and opposite to the flame acceleration in the laboratory frame.

According to what was said in the Introduction, the channel walls will be assumed ideal in that the wall friction can be neglected. However, as the analysis of Part~I has shown, heat losses to the tube walls are not negligible even in wide tubes, especially for flames near the limits of inflammability. These losses are taken into account by a correction factor $(1-\delta)$ in the flame propagation speed, where
\begin{eqnarray}\label{loss}
\delta = \frac{2K}{\rho c_p U_f}\,,
\end{eqnarray}
\noindent with $c_p$ an average specific heat capacity of fresh gas, $\rho$ its density, $U_f$ the normal flame speed relative to the fresh mixture, and $K \approx 20\,$W/m$^2\cdot$K the heat transfer coefficient.

\subsection{The on-shell equations}\label{onshellequations}

It was demonstrated in Part~I that the steady flame propagation in tubes can be adequately described by the two-dimensional model of flame propagation in a channel of width equal to the tube diameter. We thus choose Cartesian coordinates $(x,y)$ so that the $y$-axis is along the upper channel wall, $y = - \infty$ being in the fresh gas. Then the flame front position in the noninertial frame can be described by an equation $y=f(x),$ with the origin $(0,0)$ conveniently placed at the upper front endpoint, Fig.~\ref{fig2}. As in Part I, $(w,u)$ will denote Cartesian components of the gas velocity measured in units of the normal flame speed. Similarly, distances will be measured in units of the tube diameter $d$ (channel width),  the gas density in units of the fresh gas density $\rho,$ whereas $\rho U^2_f$ will be the unit of gas pressure $p.$ Then the burnt gas density is $1/\theta,$ where $\theta$ is the gas expansion coefficient.

Equations governing quasi-steady flame acceleration are readily obtained from the equations derived in Part~I by adding to the terrestrial gravitational potential the contribution due to the horizontal gravity present in the noninertial frame of reference introduced in Sec.~\ref{formulation}. The total potential is thus
\begin{eqnarray}\label{gpotential}
\phi(x,y) = -g\left(x - \frac{1}{2}\right) - ay.
\end{eqnarray}
\noindent The constant $g/2$ is inserted in this expression for later convenience. The only place where gravity appears in the on-shell equations is the on-shell value of vorticity generated by the flame,
\begin{eqnarray}
\sigma_+ = - \frac{\theta - 1}{2\theta N}(u^2_- + w^2_-)' - \frac{\theta - 1}{\theta N}\phi'_+(x),\nonumber
\end{eqnarray}\noindent where $N = \sqrt{1 + f'^2}\,,$ and prime denotes $x$-differentiation. One has $\phi'_+(x) = -g - af'(x),$ therefore, equations for the functions $w_-(x),u_-(x)$ and $f(x)$ are obtained from Eqs.~(11), (22) of Part~I by replacing $g \to g + a f'(x)$:
\begin{eqnarray}\label{masteru}
\frac{d}{dx}\left[\frac{u'_-}{f'} (1 + \alpha x) - \alpha  x\, \frac{g + af'}{f'u_-}\right] + \alpha\left(\frac{u'_-}{u_-} - \frac{g + af'}{u^2_-}\right)(w_- - \alpha) = 0\,, \\
f'(x)w_-(x) = (1 - x)u_-'(x)\,,\label{masterw}
\end{eqnarray}
\noindent where $\alpha = \theta - 1.$ At last, $u_-$ and $f$ are still related as $f = U - (1 - x)u_-,$ which follows from Eq.~(\ref{masterw}) and the evolution equation
\begin{eqnarray}\label{evolutioneq}
u_- = f'(w_- + 1)\,.
\end{eqnarray}
\noindent Here $U$ denotes the flame propagation speed with respect to the fresh mixture. The boundary conditions at the channel walls also remain unchanged:
\begin{eqnarray}\label{initialcond}
u_-(0) = U, \quad w_-(0) = \frac{\alpha}{2}\,.
\end{eqnarray}
\noindent We recall that all these relations were derived in Part I under the large-slope condition $f'\gg 1,$ or equivalently, $U\gg 1,$ and that $w_-(0)$ is nonzero because the functions $w_-(x),u_-(x),$ $f(x)$ describe the true flame structure for all $x$ except small regions near the walls, where the condition $f'\gg 1$ does not hold. In these regions, the set of functions $w_-(x),u_-(x),$ $f(x)$ is constructed as the unique continuation from the channel inside of a solution of ordinary differential equations. This construction applies equally well to the system of Eqs.~(\ref{masteru})--(\ref{evolutioneq}).

\subsection{Identification of physical solutions}\label{ident}

As in the case of uniform flame movement, numerical scrutiny of Eqs.~(\ref{masteru})--(\ref{evolutioneq}) shows that for each set of parameters $a,g,\alpha >0$ there are two continuous families of solutions parameterized by $U.$ They are called type~I and type~II solutions, the latter corresponding to larger $U,$ and are characterized as follows. The longitudinal component $u$ of the on-shell fresh gas velocity monotonically increases with $x$ in both cases, while the transversal component $w$ has a pronounced maximum at $x \approx 0.2$ in the type~I solutions, and is monotonic in the type~II solutions. Accordingly, the front is convex towards the fresh gas in the latter case, and has an inflection point at $x \approx 0.2$ in the former [see Part~I, Sec.~IIB3]. Which members of the two families of solutions can be realized in practice is determined by an additional condition of homogeneity of the gas flow far downstream. Since Eqs.~(\ref{masteru}), (\ref{masterw}) are valid in the limit of vanishing gas viscosity, the gas flow can be considered ideal only in a vicinity of the flame -- in the {\it transition domain} (see Fig.~\ref{fig2}). At larger distances (in the {\it dissipation domain}), the small but finite gas viscosity comes into play and ultimately makes the flow homogeneous. The criterion for selecting solutions satisfying this condition follows from the conservation of mass, momentum and energy.

First of all, by the overall mass conservation, the flow speed far downstream is $\theta U.$ Next, to write down momentum conservation, it is useful to introduce the reduced pressure $\tilde{p} = p + \phi$ of the fresh gas, and $\tilde{p} = p + \phi/\theta,$ that of the combustion products. Then the conservation of the longitudinal momentum component in the upstream and downstream parts of the transition domain, and in the dissipation domain gives, respectively,
\begin{eqnarray}\label{momentumup}
\tilde{p}|_{y=0} + U^2 &=& \int_0^1d\eta[\tilde{p}_-(\eta) + Nu_-(\eta)]\,, \\
\int_0^1d\eta[\tilde{p}_+(\eta) + Nu_+(\eta)] &=& \tilde{p}|_{y=U} + \frac{1}{\theta}\int_0^1dx u^2(x,U)\,, \label{momentumdown}\\
\tilde{p}|_{y=U} + \frac{1}{\theta}\int_0^1dx u^2(x,U) &=& \tilde{p}|_{y=s} + \theta U^2,\label{momentumdiss}
\end{eqnarray}
\noindent where $s=s(t)$ is the $y$-coordinate of the hot end of the tube, that is the distance traveled by the flame from the moment of ignition to the current instant $t.$ In writing these relations, it is taken into account that the reduced pressure is independent of $x$ outside the transition domain (as $w=0$ there), the mass flux through the front is equal to unity, and the wall friction is negligible.

Summing up Eqs.~(\ref{momentumup})--(\ref{momentumdiss}), and using Eq.~(\ref{gpotential}) and the jump conditions at the front, $p_+ - p_- = -\alpha,$ $u_+ - u_- = \alpha/N,$ one finds
\begin{eqnarray}\label{momentumtot}
\tilde{p}|_{y=s} - \tilde{p}|_{y=0} = -\alpha U^2 + \frac{\alpha a}{\theta}\int_0^1d\eta f(\eta).
\end{eqnarray}
\noindent Since the flow is uniform upstream of the transition domain, $\tilde{p}|_{y=0}$ in this relation can be replaced by $\tilde{p}|_{y=s-L}.$ Finally, the energy conservation in the transition and dissipation domains reads
\begin{eqnarray}\label{energytr}
q_K(U) + \theta U\tilde{p}|_{y=U} &=& q_K(0) + U\tilde{p}|_{y=0} - \dot{E}\,,\\
q_K(\infty) + \theta U\tilde{p}|_{y=s} &=& q_K(U) + \theta U\tilde{p}|_{y=U} -\dot{Q}\,,
\label{energydiss}
\end{eqnarray}
\noindent where $q_K(y)$ is the kinetic energy flux through the vertical cross-section of the channel at point $y,$ $\dot{E}$ is the total rate of change of the gas internal energy in the flame front, and $\dot{Q}$ is the total rate of viscous dissipation of kinetic energy due to the flow homogenization. It is to be noted that despite their appearance, Eqs.~(\ref{momentumup})--(\ref{energydiss}) are not exact, because the vanishing of $w$ at the boundaries of the transition domain ($y=0,U$), used in their derivation, holds true only at the leading order of the large-slope expansion [as do Eqs.~(\ref{masteru})--(\ref{evolutioneq})]. With the same accuracy, the gas enthalpy is continuous at the flame front, and hence the rate of change of internal energy
$$\dot{E} = \int_0^1d\eta N[p_-(\eta) - \theta p_+(\eta)] = -\alpha\int_0^1d\eta Np_-(\eta).$$ It remains to express the integral over the right boundary of the transition domain, appearing in Eq.~(\ref{momentumdown}), in terms of the on-shell quantities. To this end, we use the flow continuity together with the Bernoulli integrals for the fresh- and burnt-gas flows in this domain (neglecting $w$ in comparison with $u$ in the same approximation):
\begin{eqnarray}
\theta N(\eta)d\eta &=& u(x(\eta),U)dx,\\
\tilde{p}_-(\eta) + \frac{u^2_-(\eta)}{2} &=& \tilde{p}|_{y=0} + \frac{U^2}{2}\,, \label{bernoulliup}\\
\theta\tilde{p}_+(\eta) + \frac{u^2_+(\eta)}{2} &=& \theta\tilde{p}|_{y=U} + \frac{u^2(x(\eta),U)}{2}\,
.\label{bernoullidown}
\end{eqnarray}
\noindent where $(x(\eta),U)$ is the intersection of the line $y=U$ with the streamline that crosses the front at $(\eta,f(\eta)),$ Fig.~\ref{fig2}.

The relations expressing conservation of the flow momentum and energy across the transition domain, Eqs.~(\ref{momentumup}), (\ref{momentumdown}) and (\ref{energytr}), become identities on account of the Bernoulli integrals. However, the requirement that the flow in the transition domain be part of an asymptotically homogeneous solution, {\it i.e.,} Eq.~(\ref{momentumdiss}) leads to the following nontrivial relation\cite{footnote2}
\begin{eqnarray}\label{criterion}
\int_0^1d\eta N u(x(\eta),U) = \frac{u^2_-(1) + U^2}{2} + \frac{\alpha}{\theta}\left(a\int_0^1d\eta f(\eta) - \frac{g}{2} - a U\right),
\end{eqnarray}
\noindent where
\begin{eqnarray}\label{uvelocity}
u(x(\eta),U) = \left\{\theta u^2_-(1) - \alpha u^2_-(\eta) - 2\alpha g(1 - \eta) - 2\alpha a [U - f(\eta)]\right\}^{1/2}
\end{eqnarray}
\noindent is the burnt gas velocity at the right boundary of the transition domain as a function of the on-shell fresh gas velocity. The rate of kinetic energy dissipation can then be found from Eq.~(\ref{energydiss})
\begin{eqnarray}\label{keloss}
\dot{Q} = \frac{\alpha\theta U^3}{2} - \alpha U\left(a\int_0^1d\eta f(\eta) + \frac{g}{2} - \frac{a U}{2}\right) - \alpha\int_0^1d\eta N\left[\frac{u^2_-(\eta)}{2} - g\eta\right].
\end{eqnarray}
\noindent

\subsection{Expression for the flame acceleration and the quasi-steady condition}\label{quasisteady}

Since the gas outflow from the tube is unrestricted, gas pressure at both tube ends is equal to the ambient pressure $p_0,$
\begin{eqnarray}\label{jet}
p = p_0, \quad y = s, s - L\,.
\end{eqnarray}
\noindent (The end-tube conditions can be more complicated in the presence of acoustic resonance,\cite{disselhorst} but that would correspond to a completely different regime of flame propagation -- the so-called vibratory movement.) These conditions lead to a relation between the flame propagation speed and its acceleration. Namely, using Eqs.~(\ref{jet}), (\ref{gpotential}) in Eq.~(\ref{momentumtot}) (with $\tilde{p}|_{y=0}$ replaced by $\tilde{p}|_{y=s-L}$) yields
\begin{eqnarray}\label{aurel}
a = \alpha U^2\left\{L-s - \frac{1}{\theta}(s-U) + \frac{\alpha}{\theta}\int_0^1d\eta f(\eta)\right\}^{-1}\,.
\end{eqnarray}
\noindent This result has a simple physical meaning. Neglecting the last term in the braces (which is less than $U/2$) in comparison with large $L$ (if $L - s$ still is large), Eq.~(\ref{aurel}) states that the hydrostatic pressure drop through the accelerated gas in the tube is equal to the difference of dynamic gas pressure at the tube ends.

Equations (\ref{masteru})--(\ref{evolutioneq}) complemented by the relation (\ref{aurel}) constitute a closed system of integro-differential equations to be solved with the initial conditions (\ref{initialcond}). This system can be easily reduced to a single equation for the front position. Physical solutions are identified as those satisfying the relation (\ref{criterion}).

Despite the fact that the above system involves an integral relation (\ref{aurel}), all statements made in Part~I regarding the general structure of solutions remain true. Indeed, for any pair of parameters $U,a,$ Eqs.~(\ref{masteru})--(\ref{evolutioneq}) still constitute a system of ordinary differential equations. It is readily checked that this system can be reduced to a single second-order equation for the function $u_-(x),$ from which $u''_-$ can be expressed as a rational function of $u_-$ and $u'_-.$ The corresponding initial conditions for this equation are obtained by substituting Eq.~(\ref{initialcond}) into Eq.~(\ref{masterw})
\begin{eqnarray}\label{initialcondu}
u_-(0) = U, \quad u'_-(0) = U\frac{\theta - 1}{\theta + 1}\,.
\end{eqnarray}
\noindent Therefore, solution to the equation for $u_-$ is unique, if exists. Next, only solutions satisfying Eq.~(\ref{aurel}) are to be left. For a given $U,$ there is at most one $a$ satisfying this relation. In fact, if there were two such $a$'s, say $a_1$ and $a_2,$ then we would had from Eq.~(\ref{aurel})
$$\left\{L-s - \frac{1}{\theta}(s-U) + \frac{\alpha}{\theta}\int_0^1d\eta f_1(\eta)\right\} = \frac{a_2}{a_1}\left\{L-s - \frac{1}{\theta}(s-U) + \frac{\alpha}{\theta}\int_0^1d\eta f_2(\eta)\right\}\,,$$ where $f_1,f_2$ are the corresponding front positions. The two numbers $L$ and $s$ would thereby be nontrivially related. This possibility is to be discarded as unphysical, because $s$ belongs to the continuum $(0,L),$ and is independent of $L.$ At last, $U$ is to be chosen so as to fulfil the condition (\ref{criterion}). Numerical analysis shows that for any location of the flame within the tube, there are at most two such $U$'s, by one of each type. An important difference from the uniform flame movement is that physical solutions of either type do not necessarily exist: as is seen from Eq.~(\ref{aurel}), $a(s)$ has a pole near the cold tube end ($s\approx \theta L/(\theta + 1)$). Moreover, it turns out that type~II solutions die out well before this point (Sec.~\ref{rapidac}).

We can now express quantitatively the quasi-steady condition formulated in Sec.~\ref{formulation}. This condition will be met if the eigenvalues $a$ and $U$ vary negligibly during the travel of the gas elements through the transition domain. Indeed, since the flow upstream this domain is uniform, it cannot bring any unsteadiness therein, whereas the flow in the dissipation domain will remain steady as long as it is such in the transition domain, independently of the form of the function $a(t),$ because the flow accelerates uniformly along the tube by virtue of its incompressibility. In the accelerated frame of reference, the gas velocity $\geqslant U$ upstream of the flame front, $\sim \theta U$ downstream, and the length of the transition domain $= U$ (in the natural units). Therefore, the transition time $\lesssim 1,$ so the quasi-steady condition requires
\begin{eqnarray}\label{qscond}
\frac{da}{dt} \ll a, \quad \frac{dU}{dt} \ll U.
\end{eqnarray}
\noindent Next, differentiating Eq.~(\ref{aurel}) gives
\begin{eqnarray}
a \gg \frac{da}{dt} = \hspace{0,1cm} && \alpha U^2\frac{\theta + 1}{\theta}\frac{ds}{dt}\left\{L - s - \frac{1}{\theta}(s-U) + \frac{\alpha}{\theta}\int_0^1d\eta f(\eta)\right\}^{-2} \nonumber \\&&
+ 2\alpha U\frac{dU}{dt}\left\{L - s - \frac{1}{\theta}(s-U) + \frac{\alpha}{\theta}\int_0^1d\eta f(\eta)\right\} = \frac{\theta + 1}{\theta}\frac{a^2 V}{\alpha U^2} +  \frac{2a}{U}\frac{dU}{dt}\,,\nonumber
\end{eqnarray}
\noindent where $V=ds/dt$ is the flame propagation speed in the laboratory frame (differentiation of the $U$-dependent contribution to the large denominator would give rise to terms of the second order in the small $du/dt,$ $da/dt$). According to the second inequality (\ref{qscond}), the last term on the right is already $\ll a.$ Therefore, the first inequality (\ref{qscond}) can be replaced with a more convenient one $aV/\alpha U^2 \ll 1$ (since in practice $\theta = 5-8,$ the factor $(\theta + 1)/\theta \approx 1$ can be omitted). Also, since $U$ (as well as $a$) is obtained as a function of $s,$ it is useful to rewrite $dU/dt$ applying the chain rule, as $VdU/ds.$ Thus, the quasi-steady conditions finally take the form
\begin{subequations}
\begin{eqnarray}\label{quasisteadyconda}
&&a \ll \alpha \frac{U^2}{V}\,,\\
&&\frac{dU}{ds} \ll \frac{U}{V}\,.\label{quasisteadycondu}
\end{eqnarray}
\end{subequations}
\noindent Evidently, they are satisfied at the initial stage of flame propagation in sufficiently long tubes, $L \gg d$ [Cf. Eq.~(\ref{aurel})]. Numerical analysis shows that type~I solutions actually fulfil these conditions reasonably well all along the tube except near its cold end. In particular, the $U$-eigenvalue gains only about $1\%-2\%$ of its initial value as the flame travels a distance $s = 0.8L.$ As to the type~II flames, they are characterized by significantly more rapid changes of $U$ and $a,$ and hence cease to be quasi-steady earlier than type~I flames.

Examination of numerical solutions further reveals that of the two conditions (\ref{quasisteadyconda}), (\ref{quasisteadycondu}) the former is of primary importance in that (\ref{quasisteadycondu}) is not violated without violating (\ref{quasisteadyconda}). The inequality (\ref{quasisteadyconda}) may therefore be regarded as {\it the} quasi-steady condition. In the ordinary units, it reads
\begin{eqnarray}\label{ordconda}
a \ll \alpha \frac{U^2 U_f}{Vd}\,.
\end{eqnarray}
\noindent We observe that it has assumed the form of a bound on the flame acceleration, rather than on its derivatives, as suggested by the general formulation of the quasi-steady condition given in Sec.~\ref{formulation}. In general, the possibility to express this condition in terms of the higher velocity derivatives rests on the freedom to go over to the noninertial reference frame, and so implies no restriction on the value of the first velocity derivative, {\it i.e.,} acceleration. Such implication emerges only after the mechanism of flame acceleration is specified by Eq.~(\ref{aurel}), which obviously relates the flame acceleration and its derivatives. This is also the reason why the second inequality (\ref{qscond}) turned out to be a consequence of the first.

Finally, using the smallness of $dU/dt$ we can relate the quantities $U,a$ to the observed flame speed. As defined, $U$ is the flame speed relative to the fresh mixture. Therefore, the flame speed measured in the laboratory frame of reference is
\begin{eqnarray}\label{flamespeed}
V = U + \int\limits_0^t d\tau a(\tau),
\end{eqnarray}
\noindent where $t$ is the time of flame travel for a distance $s(t)$ from the ignition point.

\section{Comparison with the experiment}\label{comparison}

In this section, a detailed comparison will be made of the obtained results with the experimental data given in Ref.~\cite{mason1920a} The main part of the comparison is the analysis of Fig.~\ref{fig1} which provides the most complete information about accelerated flame movement, but other results\cite{mason1920a} relevant to the flame propagation in open tubes will also be utilized.

\subsection{The range of applicability of the theory and the accuracy estimate}\label{estimate}

Account of the heat losses to the tube walls and estimation of the calculational accuracy, made for uniform flame movement and summarized in Table~I of Part~I, apply equally well to flame acceleration in open tubes. Specifically, the values of the parameter $\delta$ determining reduction of the flame propagation speed will be taken from this table or, if missing therein, calculated using Eq.~(\ref{loss}) with the values of thermodynamic parameters given in Part~I. It is to be recalled that $\delta$'s appearing in the mentioned table refer to wide tubes, $d\gtrsim 20\,$cm; in tubes with $d\lesssim 10\,$cm they are to be multiplied by $\beta = 1.45.$ The corrected parameters $U$ and $a$ are obtained from the eigenvalues of the main system of equations by multiplying on $(1-\delta)$ and $(1-\delta)^2,$ respectively (the latter follows from the fact that in the quasi-steady regime, $a\sim U^2,$ Cf.~Eq.~(\ref{aurel})). As to the relative error of calculations, it is given in the next to last column of the table $(r_m).$ Unfortunately, this error is comparatively large ($25\%-30\%$) for fast near-stoichiometric flames in the $5\,$cm diameter tube, because the high-speed condition $U\gg 1$ is poorly satisfied in such narrow tubes (for instance, $U$ is only about $2$ for a type~II flame in the stoichiometric mixture). In mixtures near the limits of inflammability, this condition is satisfied much better ({\it e.g.}, $U\approx 6.5$ for a type~I flame in the $5.4\%$ mixture), and accordingly $r_m = 10\%-15\%$.

The only new question regarding approximations made is the ideality of the tube walls. The wall friction can be neglected, if the viscous drag exerted on the fluid by the walls is small compared to the difference of dynamic gas pressure at the tube ends, which according to Eq.~(\ref{aurel}) drives flame acceleration. This question did not arise in the semi-open configuration, because whatever horizontal bulk force acts on the gas within the tube or at its open end, it is exactly compensated by the flange at the other end. Now it is required that the pressure drop caused by the viscous drag, $\Delta p,$ be much smaller than $\alpha U^2.$ Restoring ordinary units for a moment, it is convenient to express the ratio of these quantities via the friction coefficient (a function of the Reynolds number)
\begin{eqnarray}\label{lambdadef}
\lambda = \frac{\Delta p d/l}{\varrho v^2/2}\,,
\end{eqnarray}
\noindent where $\Delta p$ is the pressure drop over the distance $l,$ and $\varrho,v$ are the fluid density and its speed relative to the tube. Initially, the fresh mixture is at rest, and the drag is due to the burnt gas flow, so that substituting $\varrho = \rho/\theta,$ $v = \alpha U,$ $l=s$ yields
$$\frac{\Delta p}{\alpha\rho U^2} = \frac{\alpha\lambda}{2\theta}\frac{s}{d}\,.$$ Thus, at the initial stage of flame propagation, when the velocity change due to the flame acceleration is not significant ($V-U \lesssim U$), the wall friction is negligible, if  $s/d \ll 2\theta/\alpha\lambda.$ Since this condition is obviously satisfied for sufficiently small $s,$ the question is whether it holds for $s\sim L.$ In this respect, it is important that the obtained bound is considerably relaxed at later stages of flame propagation. In fact, as the theory predicts and the experiments confirm, in all but the limit mixtures, the flame speed increases $3$ to $5$ times $U$ already half-way along the tube, which means that the burnt gas velocity decreases by the same amount. Therefore, denoting $r = v/\alpha U$ the relative reduction of the burnt gas velocity,
$\lambda$ in the above inequality is to be replaced by $\lambda r^2,$ so that the general requirement takes the form
\begin{eqnarray}\label{nofriction}
\frac{s}{d} \ll \frac{2}{r^2\lambda}\,,
\end{eqnarray}
\noindent where $\lambda$ is to be found under the flow conditions corresponding to the given $s.$ In particular, the value of $\lambda$ essentially depends on whether the burnt gas flow is laminar or turbulent. In the latter case, the boundary layer set in downstream of the flame front rapidly fills up the tube cross section (the boundary layer thickness $\sim y$ becomes of the order of the tube diameter at distances $y = 5d-10d,$ depending on the burnt gas speed), so that $\lambda$ can be taken from the Moody's chart for the friction coefficient in an established turbulent flow with the Reynolds number ${\rm Re} = vd/\nu.$ Things are different, however, if the burnt gas flow is laminar, because the boundary layer grows much slower in this case. Indeed, its thickness is $\approx \sqrt{yd/{\rm Re}}\,,$ and if one uses the formula $\lambda = 64/{\rm Re}$ for a steady tube flow, then the condition (\ref{nofriction}) readily shows that the maximal thickness of the boundary layer ($\sqrt{sd/{\rm Re}}\,$) is small compared to the tube diameter. Therefore, the picture of a fully established laminar flow of burnt gases, characterized by the parabolic velocity profile is inadequate, and is to be replaced by the one of a thin boundary layer. The ratio of the total friction force in the latter picture to that in the former is
$$r_f\equiv \frac{\pi d \sqrt{\nu\varrho^2 v^3 s}}{8\pi\nu\varrho v s} = \sqrt{\frac{{\rm Re}}{64}\frac{d}{s}}\,.$$ Now, multiplying by $r_f$ both sides of the identity for an established laminar flow
$$\frac{\Delta p}{\alpha\rho U^2} = \frac{32\alpha r^2}{\theta{\rm Re}}\frac{s}{d}$$ (this identity is obtained by inserting the formula $\lambda = 64/{\rm Re}$ together with $r = v/\alpha U$ into the definition (\ref{lambdadef}), and making the substitutions $\varrho = \rho/\theta,$ $l=s$), the requirement of smallness of its left hand side yields
$4r^2\sqrt{s/d{\rm Re}} \ll 1,$ or
\begin{eqnarray}\label{nofriction1}
\frac{s}{d} \ll \frac{{\rm Re}}{16 r^4}\,.
\end{eqnarray}
\noindent In view of the temperature drop near the tube wall, the Reynolds number ${\rm Re} = vd/\nu$ appearing in this formula is to be taken at a temperature significantly lower than the flame temperature, which, however, is difficult to calculate accurately. Therefore, by evaluating ${\rm Re}$ at the adiabatic flame temperature, as will be done below, one underestimates the right hand side of (\ref{nofriction1}), because the thermal drop of gas viscosity is larger than that of the gas velocity $(v\sim 1/\varrho \sim T,$ $\nu \sim T^{3/2}).$ Also, it is to be noted that the viscous drag associated with the fresh gas, set into the motion as a result of acceleration, is opposite to the drag acting on the burnt gas, which further reduces the pressure change due to the wall friction. By these reasons, the theory based on the ideal wall approximation becomes practically applicable already at distances $s$ satisfying (\ref{nofriction1}) in which the sign $\ll$ is replaced with $\lesssim.$ The relative error in the value of flame acceleration calculated in this approximation is thus bounded by the square root of the ratio of the left and right hand sides of (\ref{nofriction1}).

For example, in the case of flame propagation in a $9.5\%$ methane mixture in the $5\,$cm diameter tube, the flame speed at $s=4\,$m determined from Fig.~\ref{fig1} is $V \approx 7\,$m/s, while the theory gives $U = 1.1\,$m/s at the same location, so that according to Eq.~(\ref{flamespeed}) the velocity change due to the flame acceleration is $V-U = 5.9\,$m/s. Using $\theta \approx 7.4$ one finds $v = 6.4\times 1.1\,$m/s$ - 5.9\,$m/s =$  1.14\,$m/s, and then $r = 1.14/7.04 \approx 0.16.$ With $\nu = 3\,$cm$^2$/s ($\nu \approx \nu_0\theta^{3/2},$ where $\nu_0 = 0.15\,$cm$^2$/s is the kinematic viscosity at room temperature), the Reynolds number ${\rm Re} = vd/\nu = 190,$ indicating laminar flow conditions. Therefore, substituting these figures into (\ref{nofriction1}) gives $s \ll 180\,$m, which is fairly satisfied by $s=4\,$m. Similar considerations show that other methane-air flames propagating in the tube $5\,$cm in diameter and $5\,$m long all produce laminar flows of burnt matter, and satisfy condition (\ref{nofriction1}) all along their travel. Type~II flames satisfy this condition better than those of type~I, because of their large acceleration (resulting in smaller $r$). In this connection, the following important circumstance is to be noted. The accuracy to which (\ref{nofriction1}) is satisfied is highest at the initial (small $s$) and the final (small $r$) stages of flame propagation. On the other hand, the value of flame acceleration at any given $s$ is independent of the preceding flame evolution, provided that the flame is quasi-steady at this $s.$ This is because Eqs.~(\ref{masteru})--(\ref{evolutioneq}) and (\ref{aurel}) do not involve $V$ or any other trace of the flame past, so that their solutions (in particular, the eigenvalues $a$ and $U$) depend only on the current flame position within the tube.  Therefore, even if the wall friction were not negligible at all during some stage of flame propagation, this would not prevent an accurate comparison of the theoretical and experimental curves $a(s)$ at later times.

\subsection{Acceleration of methane-air flames in the tube $d=5\,$cm, $L=5\,$m}

\subsubsection{Processing of the data in Fig.~\ref{fig1}}\label{processing}

In order to experimentally verify the functions $a(s)$ found by numerically solving the system of Eqs.~(\ref{masteru})--(\ref{evolutioneq}) and (\ref{aurel}), the curves in Fig.~\ref{fig1} were first approximated by polynomial functions of time. For this purpose, a set of points was selected on each curve (correcting an obvious misprint on the figure: the two lower marks on the ordinate axis are $0$ and $50$), then digitized by means of the DigitizeIt software,\cite{footnote3} and used to construct best polynomial fit with the help of the Maple~15 software.\cite{footnote4} Since the experimental curves in Fig.~\ref{fig1} were obtained by the screen-wire method,\cite{wheeler1914,mason1917} with the wires spaced at $0.5\,$m, it would not make much sense to collect too many basis points: the number of points in a set should not exceed twice the travel distance measured in meters. Thus, the number of points is eight for the four curves $7.1\%$ to $9.5\%,$ while for the slow flames in $6.2\%$ and $5.4\%$ mixtures this number is five and three, respectively. The maximal degree of a polynomial is a unit less than the number of points in a set.

The flame propagation speed $V$ and its acceleration $a$ are found by differentiating the obtained polynomials. As to the flame speed, the procedure just described gives reliable results, in that changing location of the basis points on the curve does not change appreciably the result, which was also verified by independent evaluation of the flame speed using a ruler. However, the second derivative of the approximating polynomials turned out to be much more sensitive to the choice of the basis points: even such small variations in their position as horizontal shifts within the curve thickness\cite{footnote5} might change the value of flame acceleration by a factor of $2-3.$ Thus, the plots of acceleration versus time generally appear irregular, having bumps and falls which move under the shifts of the basis points (the corresponding polynomials are pathological in that their coefficients are large alternating). To reduce this ambiguity, it is necessary to add some physical requirement on the functions $a(t),$ for which purpose we turn to Ref.~\cite{mason1920a} that describes the observed flame behavior as follows: ``With all but the lower-limit mixture ($5\!\cdot\!40$ per cent methane), in which the speed of flame is uniform, there is a gradual and, so far as the records can indicate, regular acceleration of speed as the flame travels from end to end of the tube.'' Actually, regularity of acceleration, or more concretely, its monotonicity, is what to be expected in the case under consideration. Directly suggested by Eq.~(\ref{aurel}), this fact is also evident without calculations: since the flame acceleration is ultimately due to allowed free outflow of the fresh mixture from the cold end of the tube, it is the larger the shorter separation of the flame from this end, whereas in the absence of acoustic excitations, this separation decreases monotonically in time. On these grounds, the approximating polynomials have been required to have a monotonically increasing second derivative with respect to time. It turns out that this requirement removes the above-mentioned ambiguity almost completely: the remaining uncertainty in the flame acceleration does not exceed $30\%$ in all instances except the latest stage of flame travel in near-stoichiometric mixtures (where flame trajectories are nearly vertical, and hence the polynomials are extremely sensitive to the shifts of the basis points). Explicit expressions of the polynomials thus obtained are collected in Table~\ref{table1} (rounding their coefficients), together with the parameters used in the theoretical computation of the flame acceleration.

\begin{table}
\begin{tabular}{c|c|ccccc}
\hline\hline
  CH$_4$,
  & \hspace{0,1cm} $P(t),$ \hspace{0,1cm}
  & \hspace{0,3cm} $\theta$ \hspace{0,3cm}
  & \hspace{0,2cm} $U_f,$ \hspace{0,2cm}
  & \hspace{0,4cm}$\delta$ \hspace{0,4cm}
  & $r_m,$
  & type
  \\
  \hspace{0,2cm} $\%$ \hspace{0,2cm}
  & \hspace{0,2cm} cm \hspace{0,2cm}
  &
  & \hspace{0,1cm} cm/s \hspace{0,1cm}
  &
  & \hspace{0,2cm} $\%$ \hspace{0,2cm}
  \\
\hline\hline
5.4 & $-0.12+29.3\,t+0.23\,t^2$ & 5.20 & 6 & 0.43 & 13 & I\\
6.2 & $9.7+32.5 \,t+5.83\,t^2+0.034\,t^3+0.082\,t^4$ & 5.73 & 12 & 0.21 & 18 & I\\
7.1 & $3.71+53.4\,t+10.56\,t^2+0.75\,t^3+8.86\,t^4-5.19\,t^5+\,t^6$ & 6.29 & 22 & 0.11 & 21 & II\\
7.75 & $13.6+54.58\,t+27.26\,t^2+2.065\,t^3+0.84\,t^4+0.424\,t^5$ & 6.77 & 28 & 0.084 & 24 & II\\
8.5 & $10.3+94.39\,t+19.1\,t^2+26.5\,t^3-28.58\,t^4+11.56\,t^5$ & 7.05 & 35 & 0.065 & 28 & II\\
9.5 & $0.49+141.3\,t+41.35\,t^2+8.46\,t^3+7.771\,t^7$ & 7.42 & 40 & 0.055 & 30 & II\\
\hline\hline
\end{tabular}
\caption{The polynomials approximating flame trajectories in Fig.~\ref{fig1} (left part of the table; $t$ is measured in seconds), and the parameters used in the theoretical computation, found as described in Part~I. $r_m$ is evaluated for the flame type specified in the last column. The polynomial for $9.5\%$ methane flame is found under the extra condition that its coefficients $\geqslant 0.$} \label{table1}
\end{table}

\subsubsection{The slowly accelerating flames $(5.4\%$ and $6.2\%$ methane$)$}

As was already noticed in the Introduction, flame acceleration in mixtures with $5.4\%$ and $6.2\%$ methane is apparently weaker than in the other cases presented in Fig.~\ref{fig1}, which is also confirmed by the results of the digital processing of flame trajectories. It turns out that this observation is in conformity with the theory, and is naturally explained by the existence of two different regimes of flame propagation. It was found in Part~I that mixtures near the limits of inflammability sustain the type~I regime of flame propagation characterized by comparatively low speeds, whereas near-stoichiometric mixtures propagate the faster type~II flames. It is to be added now that in the case of open tubes, the two types are also significantly distinct regarding the magnitude of flame acceleration: in the type~II regime, the value of $a$ is $3-10$ times larger than that in the type~I regime. Next, it was established in Part~I that transition from one regime to the other takes place in the ranges of methane concentrations $\approx 7.2\%$ to $9\%,$ and $11\%$ to $12.2\%,$ wherein flames exhibit intermittency of the propagation regimes. We can therefore expect that in open tubes, flame propagation in the $5.4\%$ and $6.2\%$ mixtures is of type~I, and this turns out to be the case indeed, as the subsequent comparison shows. Consider first the $5.4\%$ mixture. As this is very close to the lower limit of inflammability, the flame is very slow (the normal flame speed is $6\,$cm/s). In fact, the value of the heat loss parameter in this case, $\delta = 0.43,$ is close to the known maximum beyond which flame propagation is impossible.\cite{zeldo1985} As to the flame propagation speed, the digital processing gives the value $29.3\,$cm/s for the initial speed and $0.46\,$cm$/s^2$ for the flame acceleration. However, such a detailed specification exceeds experimental accuracy in the case under consideration, because Fig.~\ref{fig1} shows part of the flame trajectory with only two basis points [the screen-wires at $50\,$cm and $100\,$cm; the third point is taken the rightmost, at $t = 4\,$s, as it is supposed to be based on the data at larger times (taking points to the left of that at $50\,$cm is meaningless, as the trajectory therein is an extrapolation anyway)]. By this reason, it is more correct to speak about the average flame speed on this interval, which is $30.2\,$cm. On the other hand, the theory gives, for a type~I flame, $22.7\,$cm/s for the initial flame speed, and $4.9\,$cm/s$^2$ for its mean acceleration, so that the average flame speed is $31.9\,$cm/s. At the same time, for a type~II solution, the initial flame speed, the mean acceleration and average speed are $36\,$cm/s, $12.2\,$cm/s$^2,$ and $50.5\,$cm/s, respectively. We thus conclude that the flame propagating in the $5.4\%$ mixture is of type~I indeed.

This consideration demonstrates agreement between the calculated and observed flame speeds only for the initial quarter of the tube, where experimental data is available. Although the predicted flame acceleration is fairly small, it might give rise to a noticeable change of the flame speed at later times, while according to the excerpt from Ref.~\cite{mason1920a}, quoted in Sec.~\ref{processing}, the flame speed in the $5.4\%$ mixture is uniform. This vanishing of acceleration near the limits of inflammability is most probably due to the heat losses which is a critical factor in the combustion of the limit mixtures. It leads to a gradual cooling of the burnt gases as they flow towards the tube end. As a consequence, their density increases, while the speed proportionally decreases, so that the dynamical pressure of the burnt gases reaching the tube end tends to that of the fresh gas at the other end, hence, the flame acceleration tends to zero [Cf. Eq.~(\ref{aurel})]. This effect is strongest in the limit mixtures because of the low normal flame speed (hence, large $\delta$) and low speed of flame propagation along the tube (hence, low burnt gas speed).

\begin{figure}
\centering
\includegraphics[width=0.6\textwidth]{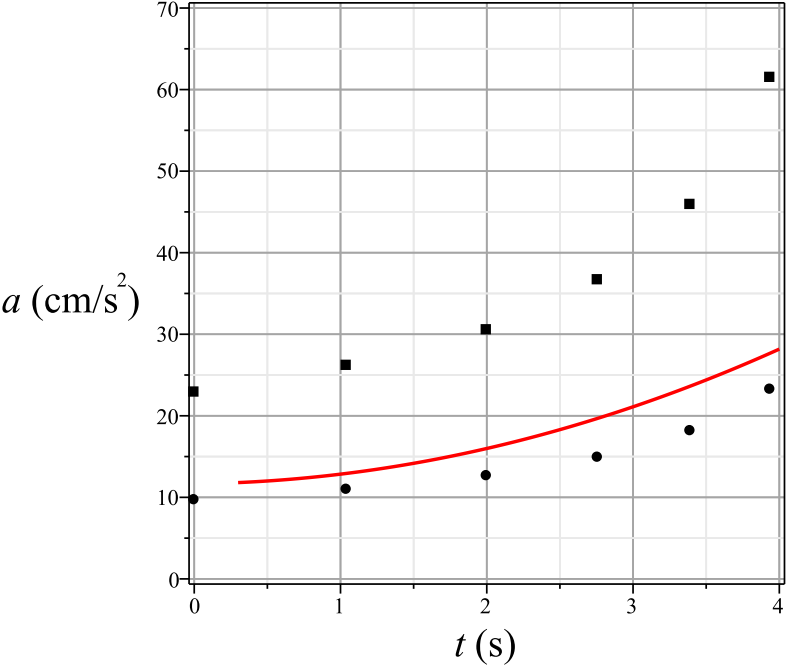}
\caption{Flame acceleration as a function of time in the $6.2\%$ methane mixture. Shown is the second derivative of the approximating polynomial from Table~\ref{table1} (solid curve), and a number of solutions to Eqs.~(\ref{masteru})--(\ref{evolutioneq}), (\ref{aurel}) [the circles and squares represent type~I and type~II solutions, respectively].}\label{fig3}
\end{figure}

Let us turn to the $6.2\%$ mixture. Here the number of basis points is five, which allows a more accurate experimental verification of the theory. The flame acceleration as a function of time is plotted in Fig.~\ref{fig3} where the solid curve is the second derivative of the approximating polynomial, and the marks are the eigenvalues of the system of Eqs.~(\ref{masteru})--(\ref{evolutioneq}), (\ref{aurel}), found for the time instants corresponding to $s = n\times 0.5\,$m, $n=0,1,...,5,$ and corrected on the heat losses as explained in Sec.~\ref{estimate}. The two sets of marks represent type~I (circles) and type~II (squares) solutions; the relative error of calculation (as quantified by the parameter $r_m$) is $18\%$ and $14\%,$ respectively, and of the same order is the experimental error. It is seen that this accuracy allows unambiguous identification of the flame as being of type~I. Thus, the flame propagates in the same regime over at least half of the tube (shown in Fig.~\ref{fig1}).

\subsubsection{The rapidly accelerating flames $(7.1\%$ to $9.5\%$ methane$)$}\label{rapidac}

\begin{figure}
\centering
\includegraphics[width=0.6\textwidth]{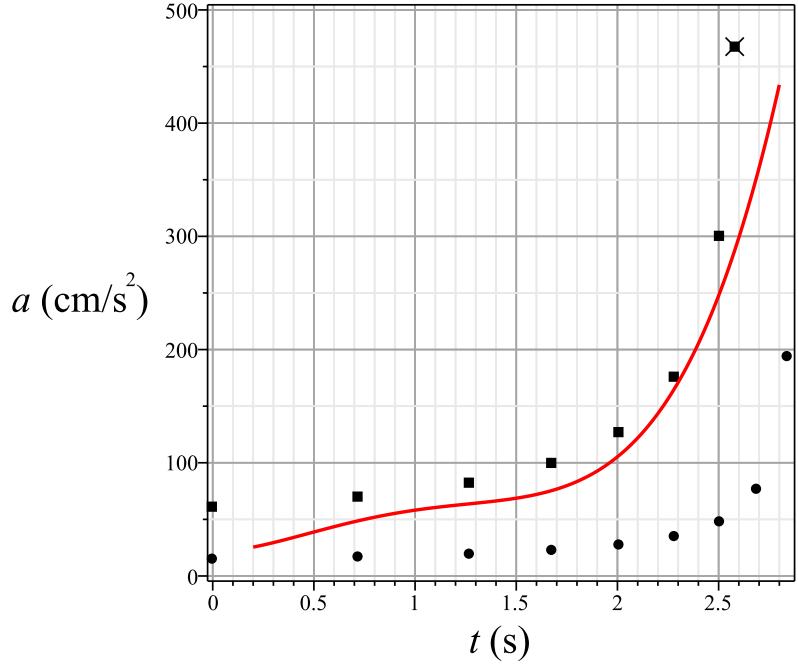}
\caption{Same in the $7.1\%$ methane mixture. The cross denotes the type~II spectrum endpoint.}\label{fig4}
\end{figure}

Consideration of the group of rapidly accelerating flames is quite similar to that just carried out, but the flame behavior is found more rich. Figure \ref{fig1} now covers a major portion of the flame travel -- $4\,$m of the total $5\,$m tube length. A new interesting element appears in the flame dynamics at distances $s \approx 3\,$m, namely, termination of the type~II regime of quasi-steady flame propagation. The exact distance where physical solutions of this type disappear depends on the mixture composition, increasing with the methane concentration, but no mixture sustains the type~II regime beyond $s\approx 4\,$m (in the tube under consideration). The spectrum endpoint is characterized by a sharp increase of the flame acceleration. It will be designated with a cross on the acceleration plots. Consider first the flame propagation in a $7.1\%$ mixture. The calculated and measured flame acceleration is plotted in Fig.~\ref{fig4}.
Although the experimental and calculational errors in this case are somewhat larger than in the preceding example (about $20\%$), the large difference of acceleration in the two regimes allows us to conclude that the flame initially propagates in the type~I regime, but then undergoes a transition to the type~II regime under which it travels a $2\,$m distance ($s\approx 1\,$m--$3\,$m). That the flame propagation in the $7.1\%$ mixture is initially of type~I is in agreement with the conclusions of Part~I, according to which in the absence of acceleration, intermittency begins only at $7.2\%$ methane concentration. Thus, the flame acceleration triggers transition to the faster regime which, however, cannot hold till the end of the flame travel because of the spectrum termination occurring at $s \approx 3.2\,$m. Beyond this point, the flame ceases to be quasi-steady, which is reflected in Fig.~\ref{fig4} showing that during the last stage of its travel, the flame remains far from the type~I regime, the only quasi-steady mode left. The question as to whether the flame might eventually return to this regime does not seem to be practically important in view of the vibratory flame movement that sooner or later develops in open tubes (as well as in semi-open) as a result of the flame-acoustic interaction.

It is to be noted that the type~II solutions in the case under consideration remain quasi-steady until they die out, despite the large value of acceleration attained: even at the spectrum endpoint, the ratio of the left and right hand sides of (\ref{ordconda}) is about $0.4.$ However, accuracy of the quasi-steady approximation goes down with growing flame speed, and in the near-stoichiometric mixtures it breaks completely well before the spectrum termination.

\begin{figure}
\centering
\includegraphics[width=0.6\textwidth]{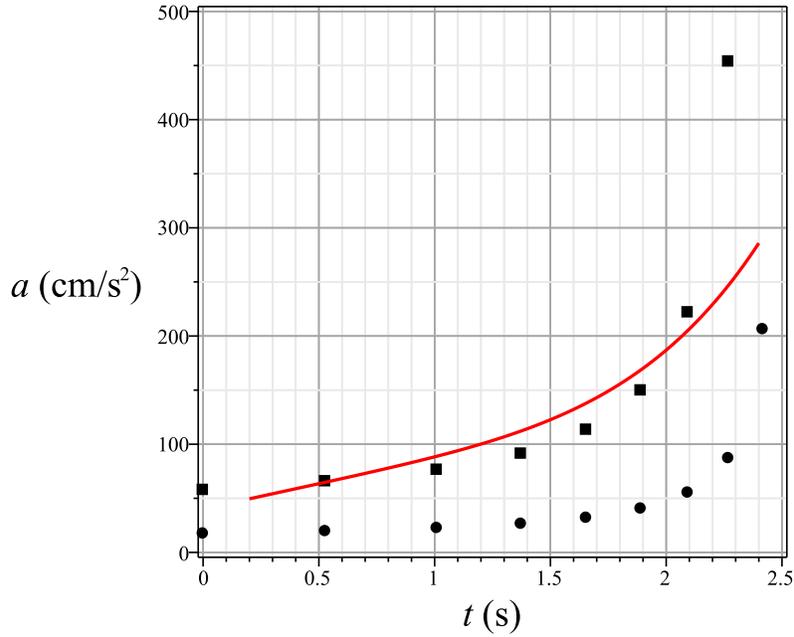}
\caption{Same in the $7.75\%$ methane mixture.}\label{fig5}
\end{figure}

Going over to the $7.75\%$ mixture, Fig.~\ref{fig5} shows that the flame evolution in this case is qualitatively quite similar to that in the $7.1\%$ mixture, except its initial stage where no trace of the type~I regime is seen. Thus, after the short transient period following ignition, the flame comes straight to the type~II regime under which it propagates until physical solutions of this type disappear. The type~II spectrum ends at $s \approx 3.6\,$m which corresponds to $t=2.3\,$s, and where the flame acceleration reaches the value $6\,$m/s$^2$ (this point is not shown on the figure in order to better resolve its lower part).

\begin{figure}
\includegraphics[width=0.6\textwidth]{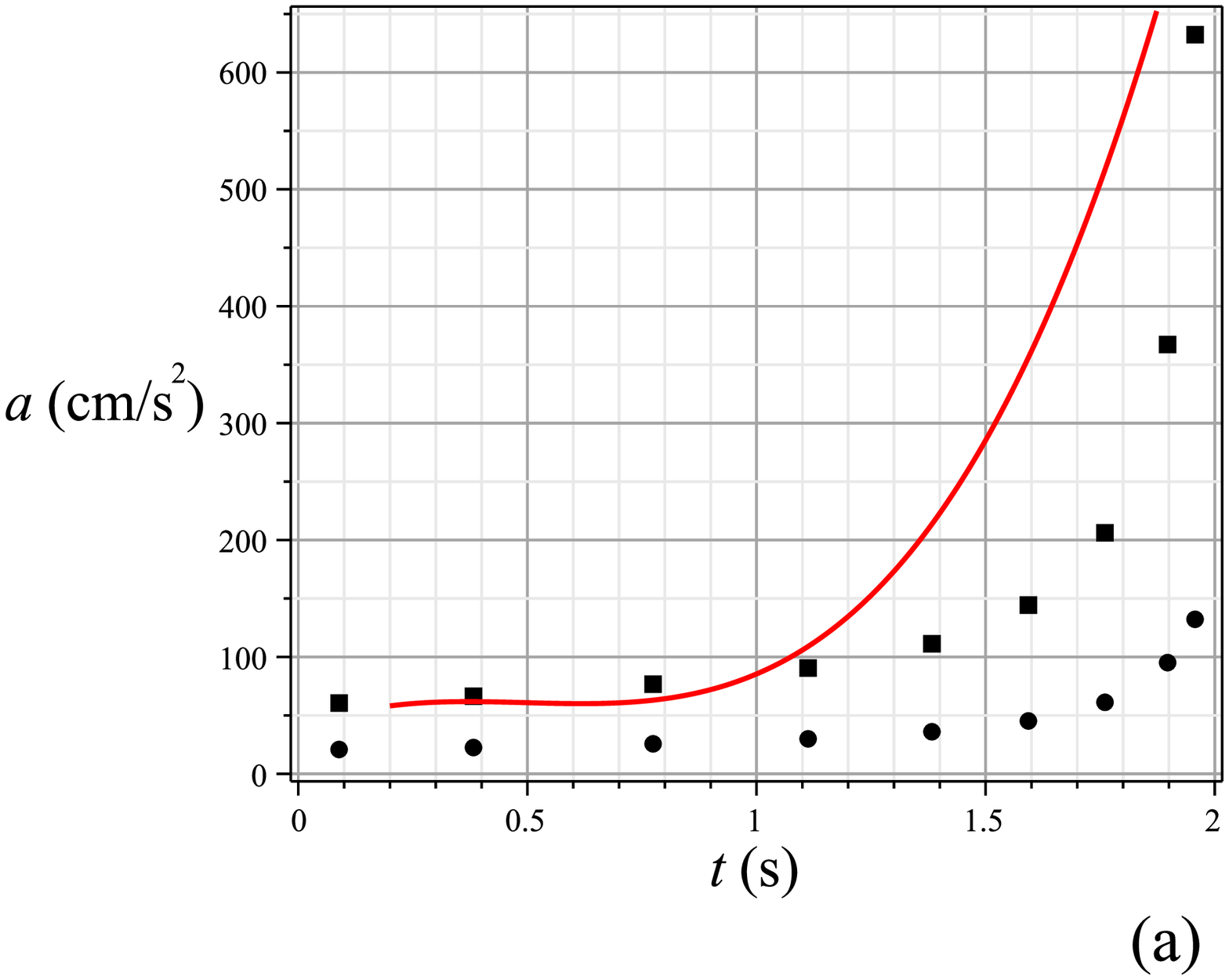}
\vskip1cm
\includegraphics[width=0.6\textwidth]{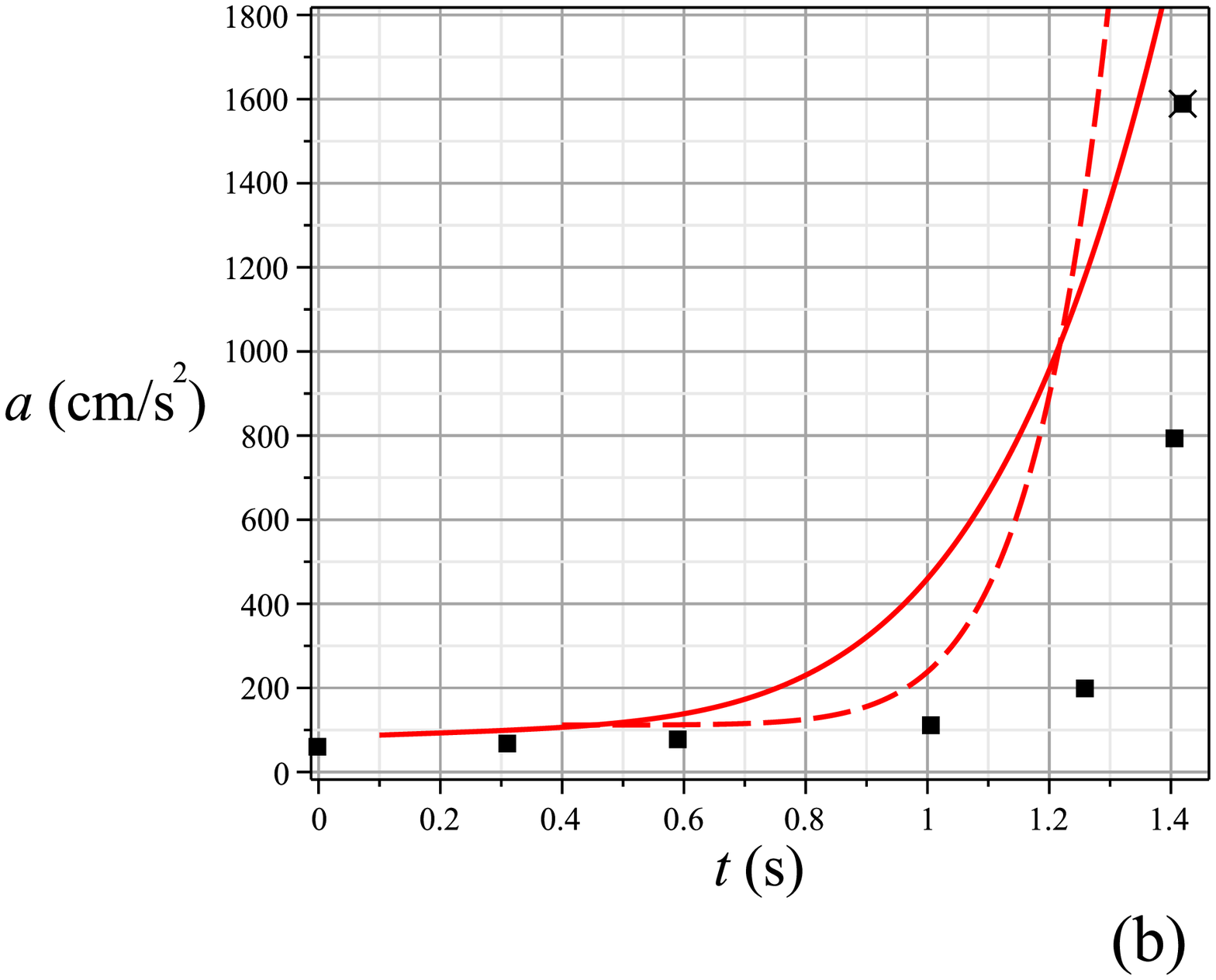}
\caption{Same in the $8.5\%$ (a) and $9.5\%$ (b) mixtures. In the latter case, only type~II solutions are shown; the dashed curve represents flame acceleration calculated using a twelfth-order polynomial with non-negative coefficients.}\label{fig6}
\end{figure}

At last, the plots of flame acceleration in the near-stoichiometric mixtures ($8.5\%$ and $9.5\%$ methane) are shown in Fig.~\ref{fig6}. The theoretical and experimental errors are largest in this case, so that the comparison can be only qualitative. The experimental error is large because of the steepness of the flame trajectories. In particular, the slope of the $9.5\%$ curve in its upper part is so large that it turned out to be necessary to impose a positivity requirement on each coefficient of the approximating polynomial in order to obtain a monotonic acceleration (without this requirement the coefficients take on extremely large values and alternate in sign). Incidentally, the form of the obtained polynomial (see Table~\ref{table1}) indicates that the polynomial approximation (under the positivity requirement on each coefficient) is probably insufficient in this case. In fact, trying various polynomials shows that increasing the polynomial degree improves the fit, but the polynomials become degenerate in that only the highest-order coefficient turns out to be nonzero together with a few coefficients of the lowest orders (describing the initial part of the flame trajectory). The plots of flame acceleration calculated using such polynomials, one of which is drawn in Fig.~\ref{fig6}(b), tend to be angle-shaped: they are relatively flat on most of the time interval, becoming very steep near its right end. As to the theoretical error, it is $30\%$ as given by the parameter $r_m$ quantifying accuracy of the steady-state calculation, but the main  error in the case of near-stoichiometric flames comes from violation of the quasi-steady condition at the later stage of flame travel. Namely, this condition is not satisfied at all near the spectrum endpoint. Specifically, for the $9.5\%$ flame, the left hand side of (\ref{ordconda}) is twice as large as its right hand side therein, so that the quasi-steady condition is violated well before the spectrum termination. Thus, the only reliable conclusion that can be drawn from Fig.~\ref{fig6} is that at the early stage of their travel, the near-stoichiometric flames accelerate in the type~II regime, as expected.

\subsection{Propagation of methane-air flames in tubes of diameters $9\,$cm and $30.5\,$cm}

Regarding flame propagation in open tubes, Ref.~\cite{mason1920a} also reports the speed measurements of methane-air flames in a $9\,$cm diameter $6.2\,$m long tube, and in tubes of diameter $30.5\,$cm. The corresponding data is discrete in that it gives the flame speed only near the tube ends, or the speed averaged over some distance. Although it does not allow direct determination of the flame acceleration, this data provides a useful check on a combination of the eigenvalues $a$ and $U.$ Denote $U_0$ the value of $U$ at the beginning of flame movement after ignition. This is the initial speed of flame propagation (after the quasi-steady regime has set in). Let the flame travel a distance $\Delta s$ (say, between successive screen-wires). If its acceleration varies insignificantly, the flame speed $V_1$ at the end of this interval can be found using the elementary formula $V_1 = \sqrt{U^2_0 + 2a\Delta s},$ while the average speed on an interval is the arithmetic mean of the initial and final speeds, $\bar{V} = (V_1 + V_2)/2.$ Consider first flame propagation in a tube with $d=9\,$cm, $L=6.2\,$m. The measured average flame speed on the interval $s=50\,$cm--$100\,$cm is given in Figure~2 of Ref.~\cite{mason1920a} For the fastest flame ($10\%$ methane), for instance, this speed is $\bar{V} = 1.67\,$m/s. On the other hand, the theory gives $U_0 = 0.93\,$m/s and $a\approx 0.96\,$m/s$^2$ for the type~II flame acceleration on the interval $s=0\,$cm--$50\,$cm ($a$ is calculated in the middle of the interval). Therefore, the flame speed at the end of this interval is $V_1 = 1.35\,$m/s. Next, the theoretical value of flame acceleration on the interval $s=50\,$cm--$100\,$cm is $a\approx 1.07\,$m/s$^2,$ hence, the final flame speed on this interval $V_2 = 1.7\,$m/s, so that the average speed therein $\bar{V} = 1.53\,$m/s. The error of calculation in this case is $25\%.$ As usual, the calculational accuracy improves for slower flames. Specifically, $r_m=15\%$ for the type~I flame in a $6\%$ mixture, where the theory gives $\bar{V} = 0.62\,$m/s while the measured flame speed is $0.57\,$m/s. It is seen that within the error of calculation, the theory agrees with the experiment, and a similar consideration shows that this is true of other cases.

Since the flames propagating in open tubes are observed continuously accelerating, the authors\cite{mason1920a} carried out experiments with a long tube ($L = 15.25\,$m) to test the possibility of a deflagration-to-detonation transition. The result was negative, and the tube length was therefore increased to $90\,$m (!). Again, no detonation occurred, but a drastic change of flame behavior at the early stage of propagation was observed: ``Instead of increasing rapidly in speed from the beginning, as when the tube was $15.25\,$m in length, the flames now travelled from the point of ignition at a constant and comparatively slow speed over a distance of between $12$ and $15\,$m (dependent on the composition of the mixture) and then began to vibrate.'' From the point of view of Eq.~(\ref{aurel}), the detected drop of acceleration is not a surprise -- the sixfold increase of the tube length reduces in the same proportion the value of $a.$ Yet, the flame speed was described as constant, though without specifying its values along the tube. At the same time, the authors emphasize (in a subsequent discussion) that this speed was somewhat larger than the speed of uniform flame movement in tubes of the same diameter. This apparent discordance can be explained as follows. Let us find the average flame speed over the initial section of the tube. As the paper gives only records for the fastest flames ($9.6\%-10.10\%$ methane), it will suffice to determine the average speed in a $10\%$ mixture. Since the tube diameter $d=30.5\,$cm is considerably larger than in the cases studied before, let us first check if the wall friction is still negligible. The theory gives $U_0=1.67\,$m/s for the initial speed of a type~II flame, so the burnt gas velocity is initially $v=6.48\times 1.67\,$m/s $\approx 11\,$m/s, and hence ${\rm Re} \approx 11200,$ indicating turbulent flow conditions. The corresponding friction coefficient $\lambda \approx 0.03$ is significantly larger than it would be in a laminar flow. Yet, substitution into (\ref{nofriction}) yields $s/d \ll 70,$ a condition which fairly holds at distances of a few meters, and where the theory gives $a = 21\,$cm/s. This value is as small as that for a type~I flame propagating in a $6.2\%$ mixture in the $5\,$cm diameter tube. As to the quasi-steady condition, in a tube with such a large aspect ratio ($L/d=295$), it holds perfectly at all distances of interest. Thus, the average flame speed over the distance $s=12\,$m is $\bar{V} = 2.2\,$m/s. Barring the wall friction that might become noticeable at the end of this distance, the error of this calculation is $20\%.$ On the other hand, the experiment gives $V = 2\pm0.1\,$m/s for the flame speed in the initial tube section, and $1.7\,$m/s for the speed of uniform flame movement. Thus, it appears likely that the small flame acceleration was not experimentally resolved (probably because of the increased screen-wire spacing), and what was specified as the constant speed of the flame was actually its speed averaged over the initial section of the tube.

\section{Discussion and conclusions}\label{conclusions}

The analysis carried out above leads us to the conclusion that in long open tubes with smooth walls, most of the flame travel proceeds in the quasi-steady regime, the difference of gas dynamic pressure at the tube ends being the reason of flame acceleration. Two essentially different regimes of quasi-steady flame propagation generally exist: in the type~I regime realized in mixtures close to the limits of inflammability, the flame speed and its acceleration are significantly lower than in the type~II regime obeyed by the near-stoichiometric flames (the ratio of flame acceleration in the two regimes ranges from about $3$ at the early stage to about $10$ near the endpoint of the type~II spectrum). The results of Sec.~\ref{rapidac} suggest that in between of these two extremes (in transient mixtures), flames begin their propagation in the type~I regime, and then undergo a transition to the type~II regime (as in the case of $7.1\%$ mixture). This observation seems to find experimental confirmation in the following excerpt from Ref.~\cite{mason1920a}: ``in most of the experiments slight vibrations were noticed at different stages in the development of the propagation, the incidence of these vibrations being earlier the more rapid was the flame.'' As these vibrations are evidently not of the acoustic origin (the vibratory movement due to the flame-acoustic interaction, once set in, continues with growing amplitude till the end of the flame travel), they presumably are a sign of the mentioned transitions between the two propagation regimes. It is to be noted in this connection that the termination of the type~II spectrum also must be observed as a spontaneous flame disturbance (if the flame propagates in the type~II regime). The spectrum terminates in the second half of the tube, at a distance increasing with the normal flame speed (Cf. Sec.~\ref{rapidac}).

As to the distance where transition from type~I to type~II regime of flame propagation takes place, this is a question of their relative stability, so that only general statements can be made in advance of a detailed stability analysis. It was argued in Part~I that in the case of uniform flame movement, the relevant dimensionless number is $\widetilde{\rm Re} = U^3_f/\nu g$ (with $\nu$ the fresh gas viscosity), a change of the flame type occurring at $\widetilde{\rm Re} \approx 180.$ Opening the tube adds a new dimensional parameter -- the tube length. The distance $s$ also is to be included into the list of independent parameters [Cf. Eq.~(\ref{aurel})]. Of these, a new dimensionless number can be formed -- $s/L$ (on the other hand, the ratio $d/L$ is to be excluded, because all the above consideration is confined to the case $d/L\ll 1,$ and the limit $d/L\to 0$ obviously exists). Thus, the transition criterion has the form
$$\Phi\left(\widetilde{\rm Re}\,,s/L\right)\approx 180,$$  
where the unknown function $\Phi(x,y)$ is such that $\Phi(x,0)=x.$ Therefore, for $s/L$ small enough, this criterion can be written as
\begin{eqnarray}\label{trcriterion}
\widetilde{\rm Re} + \phi(180)s/L \approx 180,
\end{eqnarray}
\noindent where $\phi(x)\equiv \partial \Phi(x,0)/\partial y.$ The coefficient $\phi(180)$ is positive, since increasing $L$ can only delay the transition. We conclude that the larger the normal flame speed, the closer to the ignition point it undergoes a transition from type~I to type~II regime of propagation. Of course, this simple reasoning applies only to flames close to the transition (that is, flames in transient mixtures), where the expansion of $\Phi$ and the replacement $\phi(\widetilde{\rm Re})\to \phi(180)$ are legitimate. Yet, it gives considerable support to the hypothesis that the flame vibrations described in the excerpt quoted above represent transitions between the two regimes of flame propagation. The value of $\phi(180)$ can be estimated from the experimental data for the flame in a $7.1\%$ mixture, Sec.~\ref{rapidac}, in which case $\widetilde{\rm Re} = 72,$ and the transition takes place at $s\approx 1\,$m ($s/L\approx 0.2$), hence, $\phi(180)\approx 540.$ Using this, it  readily follows from Eq.~(\ref{trcriterion}) that the $7.75\%$ flame ($\widetilde{\rm Re} = 150$) ought to undergo a transition at $s\approx 30\,$cm, that is before the first basis point ($s=50\,$cm), in agreement with the observation made in Sec.~\ref{rapidac}.

Next, several technical remarks are in order. The first concerns dependence of the solutions on the gravity acceleration. As discussed in Part~I, each type of solutions describing uniform flame movement obeys a simple scaling law with respect to the dimensionless parameter $g,$ namely, $f'\sim \sqrt{g},$ $u_-\sim \sqrt{g},$ $w_-\sim g^0.$ In other words, obtained in the leading order of the large-slope expansion, these solutions represent at the same time the large-$g$ asymptotics of the exact solutions. This is no longer so in open tubes. While Eq.~(\ref{masteru}) suggests that the flame acceleration ought to scale as $a\sim \sqrt{g},$ Eq.~(\ref{aurel}) shows that $a\sim g$ (barring the small $U$-dependent terms in the denominator), because $U=u_-(0)\sim \sqrt{g}.$ Thus, the large-$g$ asymptotic does not exist in this case. Since $g = (d/U^2_f)\times 9.81\,$m/s$^2$, this implies that the large-$d$ or small-$U_f$ asymptotics neither exist. The latter means in turn that the phenomenon of flame acceleration in open horizontal tubes in no way can be described within the bubble model ($U_f=0$).

Second, regarding the termination point of the type~II regime, it should be emphasized that the existence of this point is an intrinsic property of the type~II solutions, which is unrelated to either the singularity at $s\approx \theta L/(\theta+1)$ of the function $a(s)$ given by Eq.~(\ref{aurel}), or to a possible violation of the quasi-steady condition. This feature provides a mechanism for further increase of flame acceleration, in the following way. It is natural to expect that the flame unsteadiness occurring beyond the spectrum endpoint will increase the front curvature, and hence effectively the normal flame speed. Since the increase of the normal flame speed shifts the spectrum endpoint towards the cold tube end, the flame will travel for some distance more as a type~II flame obtained from the true flame by averaging over the unsteady disturbance, and which propagates with the increased normal speed and hence, increased acceleration. This qualitative description becomes rigorous when the characteristic length of the flame disturbance is much smaller than the tube diameter (Cf. Sec.~III of Part~I).

Finally, to complete the portraits of the two types of flames, it is to be added that they are also notably distinct regarding the value of the kinetic energy losses taking place in the dissipation domain and accompanying homogenization of the burnt gas flow. The rate of this process is given by Eq.~(\ref{keloss}). Numerical results show that this rate in the type~II regime is 1--2 orders of magnitude larger than that in the type~I regime. Moreover, the directions of its change in the two cases are opposite: the dissipation rate in the flow generated by a type~II flame appreciably increases as it travels along the tube, and somewhat decreases for a type~I flame.

\acknowledgments{This study was partially supported by RFBR, research project No. 13-02-91054~a.}


\begin{thebibliography}{}
\bibitem{footnote2}
At first sight, Eq.~(\ref{momentumdiss}) does not seem to represent an independent relation, as it brings in a new unknown $\tilde{p}|_{y=s}.$ It should be kept in mind, however, that the velocity fields depend only on the relative pressure, in particular, one can set $\tilde{p}|_{y=s}=0$ (in fact, it is not difficult to see that the conservation laws can be rewritten in the form involving only the pressure differences).
\bibitem{footnote5}
Accuracy of the screen-wire method is limited by the switch time of the Wheeler's ``commutator,''\cite{wheeler1914} which is $0.01\,$s. Inspection of Fig.~\ref{fig1} shows that this value is just the halfwidth of the curves representing flame trajectories.
\end{thebibliography}
\end{document}